# Energy and Charge Transfer for Na$^+$ Ions Scattered from a Ag(001) Surface

M.P. Ray[1], R.E. Lake[1,2], J.B. Marston[3] and C.E. Sosolik[1*]

[1]Department of Physics and Astronomy, Clemson University, Clemson, South Carolina 29634, USA

[2] Present address: COMP Centre of Excellence, Department of Applied Physics, Aalto University, P.O. Box 14100, 00076 Aalto, Finland

[3]Department of Physics, Box 1843, Brown University, Providence, Rhode Island 02912, USA

[*] Corresponding author:  Email:  sosolik@clemson.edu, Phone: 864-656-0310

**Abstract:** We present energy- and charge-resolved measurements of low and hyperthermal energy Na$^+$ ions scattered from a Ag(001) surface.  With the primary ion beam oriented along the [110] crystal direction, distinct peaks in the energy distributions of the scattered beam flux that correspond to single or multiple collisions with target atoms are observed.  A classical trajectory simulation reveals that these collisions can occur either at the surface or within the [110] channels, depending on incident beam energy.  Within the simulation we probe the role of finite temperature and thermally displaced atoms on specific scattering events and show that contributions to the scattered distributions from single and double collisions dominate within the [110] channels.  We also report velocity dependent measurements of the neutral/ion ratio of the scattered beam flux.  A deviation between the data and simulated charge transfer results is observed for Na trajectories which penetrate the surface.

1. Introduction

Studies that utilize hyperthermal to low energy (≈ 1 eV to 1 keV) ions to probe the fundamental dynamics of energy and charge transfer at surfaces are unique as they bridge the gap between adsorption-dominated thermal energy (< 1 eV) beam effects and the collision-dominated phenomena observed in the low-to-medium energy regimes (> 1keV).  The small de Broglie wavelengths of the incident projectiles ensures that scattering trajectories are inherently classical and involve only a few atoms at the surface.  This allows those trajectories to be modeled through either classical kinematic theory or small-scale molecular dynamics simulations [1-2].  Accurate computational reproductions of ion scattering data within such simulations requires species-specific scattering potentials [1,3-8]. Additionally, the intrinsic timescales involved in ion-surface interactions in this energy range allow for velocity-resolved measurements of charge state evolution as a function of atomic position outside of the target surface [1,9-13].

While numerous studies have focused on hyperthermal and low energy alkali ions scattered from the surface of Cu(001) (see [1] and references therein), here we have chosen to examine a Ag(001) target. Energy-, velocity-, and angle-resolved spectra for $Na^+$ ions with incident energies between 20 eV and 1 keV have been reproduced using a classical scattering simulation.  As our data show, the Ag(001) target, which has an increased lattice spacing relative to Cu(001), allows the incident $Na^+$ ions to penetrate the surface and scatter from second layer Ag atoms.  Therefore, these results explore the transition from purely ion-surface interactions to ion-solid interactions, where the ion can interact directly with below-surface atoms.  The resulting subsurface trajectories are well-suited to test current resonant charge transfer theory which treats the neutralization of a scattering ion as independent of its trajectory and dependent only on the scattered velocity.  We compare our charge-resolved measurements to a quantum mechanical model which is able to reproduce the velocity-dependent trends for all data except for those velocities which correlate with ions that have penetrated the surface.  The origin of this deviation is discussed relative to the presence of subsurface trajectories and the assumptions of our charge transfer model.  These results are of general interest as there is a continued focus on the fundamental nature of charge exchange for simple s-shell systems such as $Na^+$ at surfaces [11-12,14].

The organization of this paper is as follows. In Sec. 2 we describe our experimental apparatus as well as the specific methods used to obtain both scattered spectra and charge transfer data from $Na^+$ beams on Ag(001). The results of these two types of measurements are discussed in Sec. 3.  Additionally, we compare our data to both a classical trajectory simulation and the quantum mechanical model of the atom-surface charge transfer problem. The full results as well as prospects for future measurements are summarized and discussed in Sec. 4.

2. Experiment

The instrument used to conduct the measurements described here is a UHV hyperthermal and low energy (< 5eV to 10 keV) ion beamline capable of producing mass-resolved, monoenergetic beams of alkali and noble gas ions.  Attached to the beamline is a two-tier UHV scattering chamber with a working pressure of 5 x $10^{-10}$ Torr.  The upper tier of the scattering chamber houses several surface analysis tools including a low energy electron diffraction (LEED) system and an Auger electron spectrometer.  The lower chamber tier contains a 180° electrostatic analyzer (ESA) for detecting scattered ions and a neutral particle detector (NPD) for detecting scattered neutral particles.  A manipulator stage is used to translate substrates between the two chamber tiers for analysis and scattering experiments.  The full instrument has been described in detail elsewhere [15-19].

The target substrate for these ion beam measurements was a Ag(001) single crystal [20]. The crystal was oriented so that the beam was incident along the [110] direction as verified using a combination of

LEED and ion scattering spectroscopy. The standard experimental procedure for both the ESA and NPD measurements involved first a substrate cleaning, which consisted of repeated cycles of sputtering with 500 eV Ar$^+$ ions followed by an anneal to 425$^o$C. After cleaning, the Ag(001) sample was allowed to cool to room temperature. The sample was then moved into the path of the Na$^+$ beam for ESA and NPD data collection. All NPD data were collected first in any given data set. After each beam exposure, an Auger electron spectroscopy scan was performed to ensure that no surface contamination was present. It is known that prolonged exposure of the target surface to an alkali beam can alter rates of neutralization [1]. However, no such changes were observed in repeated measurements, and the aforementioned cleaning procedure served to maintain a well-defined and clean target across the multiple data sets presented here.

The ESA was used to obtain angle- and energy-resolved spectra of scattered Na$^+$ ions from our Ag(001) crystal for incident energies ranging from 20 eV to 2 keV. The ESA has an energy resolution of $\Delta E/E=0.016$ and an effective angular acceptance of approximately ±0.5$^o$. Spectra were measured with the incident beam fixed at either 45$^o$ or 55$^o$ relative to the surface normal. Both the scattering geometry and a sample spectrum are shown in Fig. 1. The distinct peaks present in this $\theta_i= \theta_f =45^o$ spectrum correspond to specific ion trajectories at the surface. Similar energy-resolved peaks are seen in the $\theta_i= \theta_f =55^o$ spectra of Fig. 3. These data are compared to the results of a classical trajectory simulation in Sec. 3.

The NPD was used to obtain velocity-resolved neutralization probabilities for studying charge transfer in the Na$^+$-Ag(001) system. Using this detector and a combination of beam pulsing and standard time-of-flight techniques, the total scattered flux (ions and neutrals) and the neutrals-only scattered flux were measured [21]. A representative data set is shown in Fig. 2, where the beam was scattered specularly at 55$^o$ from the Ag(001) surface. At each incident beam energy, the total scattered flux and neutrals-only scattered flux spectra were integrated. The ratio of these integrated intensities was taken as a measure of the neutralization probability at that incident energy ($P_0(E_0)$).

## 3. Results and Discussion

Both scattering and neutralization data were obtained using the methods described in the previous section. The incident energy of the Na$^+$ beam was varied between 20 eV and 1000 eV for both 45$^o$ and 55$^o$ incidence angles. Our scattering data were compared to the simple predictions of the sequential binary collision approximation (SBCA) as well as to results of the classical trajectory simulation SAFARI [22]. The neutralization data were compared to the results of a quantum mechanical simulation [23].

### *3.1 Binary Collision Approximation*

A simple approach that can be used to calculate scattered ion energies in this type of experiment involves the use of classical mechanics principles, i.e. billiard ball physics. In fact, a straightforward application of energy and momentum conservation will yield the scattered energy if the incident energy and total scattering angle are known. This approach can be generalized in the form of the kinematic factor expression given as:

$$k_\pm(\mu, \theta_{TSA}) = \frac{E_f}{E_0} = \frac{\mu^2}{(1+\mu)^2}\left[\cos\theta_{TSA} \pm \left(\frac{1}{\mu^2} - \sin^2\theta_{TSA}\right)^{1/2}\right]^2$$

The kinematic factor, $k_\pm$, gives the ratio of the final and incident energies of the projectile ion, $E_f$ and $E_0$, following a single collision with a target atom in terms of the projectile-to-target mass ratio $\mu=m_{proj}/m_{target}$ and the total scattering angle, $\theta_{TSA} = 180° - \theta_{incident} - \theta_{final}$. While two possible solutions for this factor, differing by a sign, are indicated, we need only consider the $k_+$ solution for light ($\mu <1$) atom-surface scattering systems such as Na-Ag(001) [25]. This kinematic factor can not only be applied in cases where the ion scatters from a single target atom but can also be applied sequentially to determine the scattered ion energy for cases involving more than one collision.  We note that for small total scattering angles, the kinematic factor expression tends toward its maximum value as one would expect for such glancing-type collisions.  Therefore, scattered trajectories which involve multiple glancing collisions can return final energies (the product of multiple kinematic factors) that are larger than a trajectory involving a single collision event which involves a single large total scattering angle.  As we see below, this occurs for Na-Ag in the case of a double collision event.  This approach is generally referred to as the sequential binary collision approximation (SBCA) and for the collision energies used in this experiment, numerous spectra have been obtained in other measurements which show features indicative of single, successive, and collective ion-atom collision events [see, e.g. 8,24].

Figure 1 shows the kinematic factor (dashed line) for a single collision between a Na and a Ag atom.  Also shown is the factor for a double collision (dashed-dotted line) which corresponds to a Na$^+$ ion scattering sequentially from two Ag atoms.  Between these sequential collisions, the ion travels parallel to the surface plane along the [110] direction before finally scattering into the detector [8,24].  We observe reasonable agreement between the single scattering kinematic factor and one of the most intense peaks present in the data.  Similar agreement is also seen for the double collision expression; however it consistently gives results for the scattered energy that are above all observed peaks in data taken across all final angles.  In addition, there are energy-resolved features that appear in the spectrum of Fig. 1 as well as in spectra obtained at other angles that do not conform to either the single or double collision predictions given by the SBCA.

There are also significant features that arise from other factors not included within the SBCA formalism.  In particular, the SBCA makes no allowance for the evolution of allowed/non-allowed trajectory types as the incident energy is varied.  The importance of this can be seen in the spectra of Fig. 3, where incident energies over the range of 50-2000 eV are shown with $\theta_i = \theta_f = 55°$.  The two prominent energy-resolved peaks that appear in these spectra for incident energies above 50 eV represent the single and double collisions described above.  However, it is clear that at both the highest and lowest incident energies a low energy shoulder appears near the single collision peak.  Additionally, for $E_0 \lesssim 100\ eV$, significant scattered intensity appears between the single and double collision peaks.  In order to understand these spectra in detail, we must consider the role of complex trajectories (beyond pure single and double) as well as recoil/binding of the surface atoms and the presence of an image charge in the metal target.  These effects can be included in a straightforward manner using a classical trajectory simulation as we discuss below.

*3.2 Classical Trajectory Simulation*

The classical trajectory simulation SAFARI was used to model the scattering of Na$^+$ from Ag(001).  In this simulation, Hamilton's equations of motion are integrated for an ion interacting with a surface in order to obtain angle- and energy-resolved spectra.  The simulation assumes that the energy loss for a scattered ion is due solely to momentum transfer to the recoiling surface atoms.  Although one could consider additional energy loss channels such as electron-hole pair formation, the typical losses to this channel are less than a few tenths of an eV for ions interacting with metals at our perpendicular velocities [26-29]. Such small contributions would not alter our results and are therefore not included in the present simulations [30].

This code has been used successfully in several other studies involving alkali ions and noble metals [7-8,22,24,36-39]. Using SAFARI it is possible to incorporate the factors outlined above that were not included in the SBCA.  For example, each atom in the crystal target is bound to its nearest neighbors to allow for the recoil of "bound" atoms.  Also, more complex trajectories beyond the single and double collision types that appear in the SBCA analysis appear naturally in the simulation because all possible impact points within a surface unit cell are probed. Our results show this is the most important factor for reproducing the features observed in the spectra as it allows for penetration of incident Na⁺ ions into the near surface region to yield more complex trajectories.

In order to determine the trajectory types that contribute to the scattered intensity at specific incident energies, simulated spectra were obtained within SAFARI. An adaptive grid (AG) method that iteratively samples impact parameters for incident Na⁺ ions within one Ag surface unit cell is used to select only those regions of a cell that lead to scattering into a predefined detector. From within these regions, specific impact points can be chosen for single-shot SAFARI runs to determine the ion's path. The predefined detector used in the simulation was circular with a 3° half angular acceptance and an energy resolution of Δ E/E=0.01. The increase in the angular acceptance of the simulated detector serves to reduce computation time and broaden the peak widths in the SAFARI spectra. We note that the increased angular acceptance does not significantly affect the mean energy of the peaks, the relative peak intensities, or the determination of trajectory type.

### 3.3 *Ion-Surface Interaction Potential*

A user-defined interaction potential is required to compute the forces that act between the incident Na⁺ ion and the Ag crystal within SAFARI. For this work, we have utilized a potential that combines two terms: the repulsive pair interaction, $V_{pair}(z)$, between a single Na⁺ ion and a single Ag atom and the attractive bulk interaction, $V_{attr}(z)$, that arises due to the image charge formed in the metal. More specifically, within SAFARI the full repulsive term is taken to be the sum of the individual repulsive contributions from the ten nearest Ag atoms to the Na⁺ ion, $V_{rep}(r) \sum_{i=1}^{10} V_{pair}(r_i)$. It was found that using ten nearest atoms sufficiently reproduced the experimental results while keeping computational time to a minimum. We determined the single repulsive term $V_{pair}(z)$ by first calculating the ground state energy of an isolated Na-Ag dimer with interatomic separations between 0.5 Å and 2.0 Å. The isolated ground state energies of individual Na and Ag atoms were then calculated and substracted from the dimer values to isolate the repulsive contribution to the energy, *i.e.* $V_{pair}(z)$ =E[Na-Ag]⁺(r) - E[Na] - E[Ag]. The values for $V_{pair}(z)$ were all calculated using the Hartree-Fock code in the quantum chemistry package GAUSSIAN 98 [40].

The bulk image contribution to the full interaction potential is represented within SAFARI as a *z* - dependent function

$$V_{attr}(r) = -e^2 / \sqrt{16(z - z_0)^2 + e^4/V_{min}^2}$$

where *z* is the perpendicular distance from the top layer of Ag surface atoms. Written in this way, $V_{attr}(z)$ is saturated to $V_{min}$ close to the surface and tends smoothly to 1/4z for large values of *z*. $V_{min}$ and $z_0$ determine the depth of the image well and are the only adjustable parameters in the total potential. A $V_{min}$ value of 2.0 eV was found to give the best agreement with the experimental data, and $z_0$ was taken to be 1.26 Å from the atomic cores [41].

### 3.4 *Ion Penetration Depth*

To understand the evolution of the spectra and the associated trajectories seen in the experimental data of Fig. 3, SAFARI AG results were obtained across the full incident energy range for surface temperatures of 0 K and 300 K. The contributions from individual trajectories in each spectrum were examined, and two trends were observed. First, a clear transition from top-layer to second-layer scattering trajectories appears as the incident energy is increased. Also, the relative contribution of zig-zag (ZZ) or in-surface-plane trajectories to the spectra is a strong function of the beam energy. Evidence for these trends can be seen by examining the minimum ion-surface distance ($z_{min}$) or penetration depth achieved by the incident Na$^+$ ions as a function of energy.

In Fig. 4, we plot $z_{min}$ as a function of the incident energy, where the $z_{min}$ values are taken as a weighted average of all the penetration depths for trajectories that reach the simulated detector and contribute to the specular scattered intensity. Looking first at the T=0K case, we observe a distinct non-monotonic variation in $z_{min}$ for 75 eV < $E_0$ < 175 eV. This behavior signifies the onset of the in-surface-plane ZZ trajectory types, which serve to lower the weighted $z_{min}$ values heavily towards zero (the surface plane) within a narrow energy range. This effect is lessened and $z_{min}$ increases again as the incident energy is increased and these trajectories no longer dominate the spectrum. Experimental evidence for these trajectory types is clearly seen in the 101 eV spectrum of Fig. 3, where an intermediate peak attributable to zig-zag scattering appears between the single and double collision peaks. The zero temperature $z_{min}$ values also show a clear transition from top layer ($z_{min}$ >0) to second layer ($z_{min}$ <0) scattering for incident energies greater than 400 eV. Following this transition, there is little variation in the penetration depth with increased incident energy, and we can consider the scattered spectra to be dominated by trajectories that interact with the second layer of the Ag(001) surface.

Also shown in Fig. 4 are $z_{min}$ values obtained at T=300K. First, it is clear that the elevated temperature leads to a loss of the non-monotonic trend in $z_{min}$ for 75 eV < $E_0$ < 175 eV. At T=300K the displacement of surface atoms from their zero temperature lattice positions lowers the probability that the ZZ trajectory types can occur since they rely on the presence of multiply aligned atoms in the surface plane. Therefore, with fewer ZZ trajectories contributing within this energy range, the $z_{min}$ values obtained are less weighted toward the surface plane value. Also at T=300K we observe a reduction in the threshold energy for the transition from top layer to second layer scattering. This can also be attributed to the loss of ZZ events, since their absence accelerates the drop in the weighted $z_{min}$ value towards the second layer value. Beyond this threshold for penetration into the second layer, any dependence on temperature for $z_{min}$ is lost, which indicates the dominance of second layer QS and QD events that are relatively insensitive to thermal displacements.

### 3.5 Trajectory Analysis

The analysis from the previous section and Fig. 4 show a penetration threshold at ≈ 400 eV. At this threshold there is a shift from top layer scattering to subsurface scattering. The analysis also shows we have observed many of the trajectory types identified in previous work [8,24]. As noted above, we see evidence for single and double collisions in the data. In the context of SAFARI these are replaced with the terms quasi-single (QS) and quasi-double (QD) to account for the effects of neighboring atoms on the trajectory. In addition, ZZ trajectories appear. These correspond to events that involve the incident ion being scattered into the plane of the surface and undergoing one or more collisions before re-emerging into the scattering plane defined by the surface normal and the detector position. Examples seen in previous work are the double zig-zag (DZZ), triple zig-zag (TZZ), and quadruple zig-zag (QZZ) trajectories [8,24].

The observed trajectories for incident ion energies below the penetration threshold of Fig. 4 have been observed and described in prior work [8,24]. Here, we focus on a new class of ion trajectories, namely subsurface scattering trajectories. The spectrum with the richest set of subsurface trajectories was the highest energy spectrum in Fig. 3. Analyzing all spectra shown in Fig. 3 with SAFARI reveals that the spectrum which has incident energy of 2013 eV is representative of all spectra that exhibit subsurface scattering trajectories. Therefore, the trajectory analysis here focuses on describing these subsurface trajectories. A comparison of the $E_0$=2013 eV spectra obtained experimentally and with SAFARI is shown in Fig. 5 and four features are prominent in this spectrum. We examine each of these four features in order energetically. First we note that SAFARI reasonably reproduces the relative intensities of the peaks seen within the scattered experimental spectrum. We have divided the simulated spectrum into four energetic regions: **A**($E_0$ < 1410 eV), **B**(1410 eV <$E_0$<1540 eV),**C**(1540 eV <$E_0$<1670 eV), and **D**($E_0$>1670 eV) to facilitate discussion. The contributing trajectories within each region are listed in Table 1 where the single digit following each trajectory type indicates a first layer (1) or a second layer (2) trajectory.

| Region | Trajectory Types |
|---|---|
| A | Triple Zig Zag-1, Triple Zig Zag-2 |
| B | Quasi-Single-1, Quasi-Single-2, Double Zig Zag-2, Triple Zig Zag-2 |
| C | Quasi-Double-2(steered), Double Zig Zag-2, Triple Zig Zag-2 |
| D | Quasi-Double-1, Quasi-Double-2, Double Zig Zag-2(steered) |

**Table 1:** Trajectory types contributing to the four labeled regions in Fig. 5.

From the trajectory types listed, we first note that the QS and QD trajectory types are confined to regions B-D. In particular, both first- and second-layer QS scattering occurs only in peak B. In peaks C and D we observe QD-type events, with peak D containing the main contribution from first- and second-layer QD trajectories. The second-layer QD events located in peak C are labeled ``steered'' because they interact strongly with the first-layer Ag atoms both before and after their two primary collisions within the [110] channel.

For energies less than 400 eV we have found that the ions do not penetrate below the first layer. In the case where the Na$^+$ beam is scattered specularly from the target at 55°, only two peaks are observed. The two peaks in the scattered spectrum can be described quantitatively by the binary collision approximation as the quasi-single (QS) and quasi-double (QD) trajectories. As shown in Fig. 1, at an incident angle of 45° the ions do not penetrate the surface layer. At the more normal incident angle, the ZZ trajectories appear with more prominence than at higher incident angles and energies. The combined data-SAFARI results indicate that at the more normal angle of incidence the ion probes the [110] channel producing DZZ, TZZ and QZZ top layer trajectories [31].

### *3.6 Resonant Charge Transfer*

Measurements to probe the neutralization probability of Na$^+$ scattered from Ag(001) were conducted in a specular scattering geometry with $\theta_i$= $\theta_f$ =55°. As described previously, the neutralization probability, $P_0$ was obtained by integrating and taking the ratio of the neutrals-only and totals spectra from the NPD. The values obtained for $P_0$ at incident energies between 20 eV and 2 keV are shown in Fig. 6 as a function of the final scattered perpendicular velocity. These $P_0$ data show a velocity-dependent trend, dropping from $P_0$≈70% to $P_0$≈30% with increasing scattered perpendicular velocity.

We also note that the $P_0$ values presented here do not show any measurable trajectory-dependence across any given spectrum.

The observed velocity-dependent trend arises due to the inherent distance-dependence involved in the tunneling of electrons between a Na atom and the Ag(001) surface. In the simplest physical interpretation, we note that the potential barrier for electron tunneling increases in width as a Na atom scatters and leaves the surface. Therefore we can consider the charge state of the atom "frozen in" at a separation from the surface where the transfer of charge becomes negligible. This separation distance will vary as a function of the atom's time spent near the surface, or equivalently as a function of its scattered perpendicular velocity. That is, high(low) perpendicular velocities correspond to close(far) atom-surface separations for the freezing of the charge state. The fact that the velocity-dependence manifests itself as a decrease in the neutralization probability arises from the image charge interaction discussed in Sec. III as well as the ≈ 0.7 eV difference in energy between the bare ionization potential of Na (5.14 eV) and the surface work function (4.40 eV) [42]. The image interaction makes it energetically favorable for the Na to be positively charged close to the surface, hence higher velocity Na atoms will be less likely to exhibit neutralization [1].

Typically, full theoretical treatments of the resonant charge transfer process that drives the neutralization of alkali ions at noble metal surfaces begin with the time-dependent Newns-Anderson Hamiltonian (NAH) [29]. The NAH can be written in various forms, taking into account multiple atomic levels, excited state formation, and on-site Coulomb repulsion. However, for the measurements discussed here the NAH can be used in its simplest form, where only one atomic level is included and the electron is considered to be a spinless fermion. This simple Hamiltonian, $H(t)$, is written as [29]

$$H(t) = \varepsilon_a[z(t)]n_a + \sum_k \varepsilon_k n_k + \sum_k \{V_{ak}[z(t)]c_a^\dagger c_k + V_{ak}^*[z(t)]c_k^\dagger c_a\}$$

where $\varepsilon_a$ is the atomic level energy and $\varepsilon_k$ is the energy of a metallic level with momentum $k$. The $c_a^\dagger, c_a$ and $c_k^\dagger, c_k$ are the creation and annihilation operators for the atomic and metallic levels, respectively. These operators and the coupling matrix elements between the atomic and metallic levels, $V_{ak}$, account for the tunneling of electrons between the atom and metal. The terms $n_a$ and $n_k$ are the number operators for the atomic and metallic levels and are constructed from the creation and annihilation operators. Time-dependence is included in this Hamiltonian through $z(t)$, where $z(t)=v_z t$ and $v_z$ is the atom's scattered perpendicular velocity.

The presence of tunneling in this problem, represented by the coupling matrix elements, $V_{ak}$, can be interpreted as giving a finite lifetime to the atomic state. In this context, we can assign a width, $\Delta$, to the state that is connected to the $V_{ak}$ in our Hamiltonian by the Fermi golden rule formula:

$$V_{ak}^2 = \frac{\Delta}{\pi\rho}$$

where $\rho$ is the density of states in the metal. In simulating this charge transfer problem, we assume there are $M$ discrete metallic states in a flat band of states of half-width $D$, which gives a density of states $\rho=M/D$ [23].

The solution used here to model resonant charge transfer is an independent particle simulation described in Ref. 23. This simulation relies on the simplicity of the spinless one-level NAH, which makes it possible to integrate the differential equation for $c_a(t)$ in the interaction picture. By doing so, the occupancy of the atomic level, $\langle n_a(t) \rangle = \langle c_a^\dagger(t)c_a(t) \rangle$, can be calculated. At long times, i.e. far from the surface, this is the quantity measured in an experiment, and using numerical integration, it can be obtained. A similar solution to the NAH was used in the work of Kimmel *et al.* [43-44], and good agreement with experimental results was obtained.  More recently, a Fermi-Dirac distribution function, to account for both velocity-smearing and temperature effects was derived and included to give a good comparison with thermally dependent neutralization data obtain for Na$^+$ scattered from a Cu(001) surface [12].

In Fig. 6, the neutralization of the scattered beam as a function of incident perpendicular velocity and the results from the independent particle simulation are shown.  The neutralization is consistent with resonant charge transfer as evidenced by the excellent agreement with the theoretical results.  At the highest incident velocities the data deviate from the theory.  This could be due to an overestimation of the broadening of the atomic level close to the surface, exaggerating the probability of neutralization at larger velocities.  However, we note that the model deviates from the data at velocities where subsurface trajectories begin to dominate the scattered spectra suggesting that the overestimation in the neutralization could be trajectory dependent. Trajectory dependent neutralization has been previously observed in a similar system [45] and it was discovered that a local change in the electrostatic potential caused by the displacement of atoms at the surface increased the neutralization, in contrast to the data in Fig. 6.  Thus trajectory dependent neutralization is unlikely the cause of the trends observed in our data.  Alternatively, in our charge transfer model we assume a 1/z type image potential in the calculation of the neutralization which could break down in the subsurface trajectory regime.  However, it is unclear how a breakdown of the image potential would affect the neutralization.  To fully understand how these subsurface scattering trajectories alter the neutralization along the outgoing trajectory, further theoretical and experimental work must be performed.

## 4. Conclusions

We have presented energy and charge transfer measurements for Na$^+$ scattering from a Ag(001) single crystal along the [110] direction.  The scattered energy loss spectra were interpreted using a classical scattering simulation.   Energy loss spectra for a range of incident energies were presented and a threshold for subsurface scattering was determined.  Using the classical scattering simulation SAFARI we are able to determine specific ion trajectories associated with peaks in the energy loss spectra for surface and subsurface scattering events.  In the subsurface scattering events, the observed peaks were associated with trajectories which have been previously observed.  However, in previous studies these trajectories were associated with top layer scattering and not subsurface scattering as reported here.  Also, the simulations revealed that the subsurface scattering events were strongly temperature dependent.  The neutralization of the scattered beam varied from ≈30% to ≈70% depending on the incident velocity.  The neutralization measurements were interpreted with a fully quantum mechanical model.  The qualitative trend with incident perpendicular velocity is consistent with a resonant charge transfer mechanism; however, for incident velocities where the ion penetrates the surface, a deviation between the experimental and theoretical results is observed.

**Acknowledgements**

We acknowledge financial support from the National Science Foundation under grants NSF-CHE-0548111 and NSF-DMR-0605619.

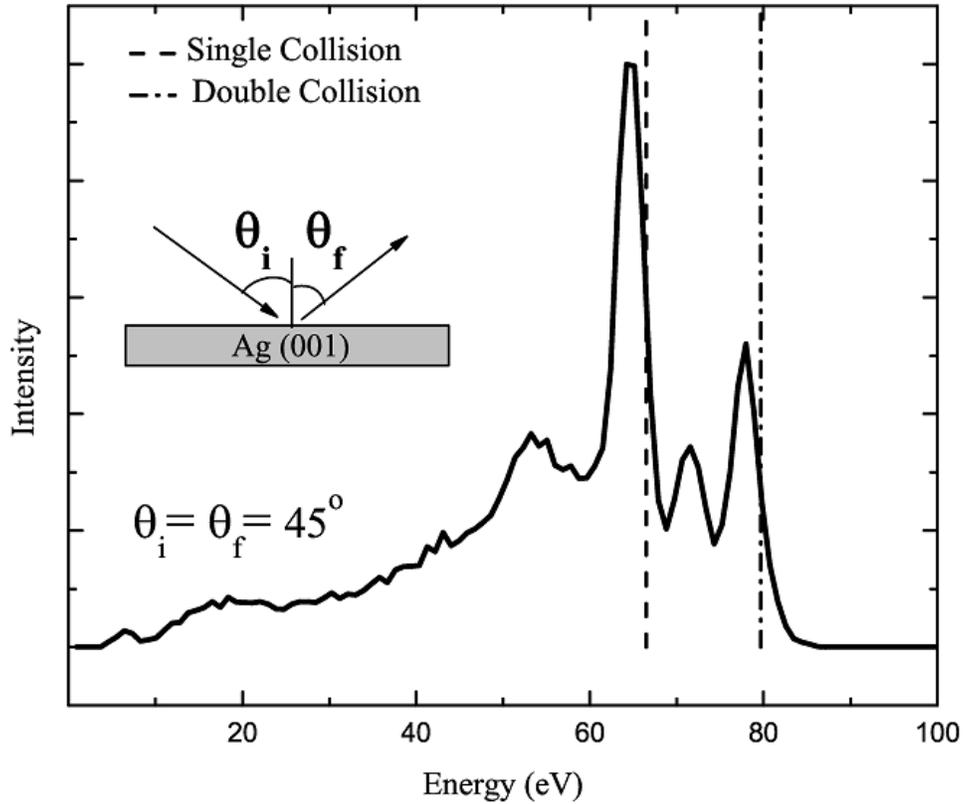

FIG. 1: A representative ESA spectrum obtained for the specular scattering of Na$^+$ from Ag(001) along the [110] direction at $E_0$ = 102.4 eV. The intensity has been normalized by 1/E to compensate for the detector's transmission function. The dashed (dashed-dotted) line shows the SBCA prediction for single (double) scattering. Note that the SBCA overshoots the observed data. The scattering geometry is shown in the upper left inset.

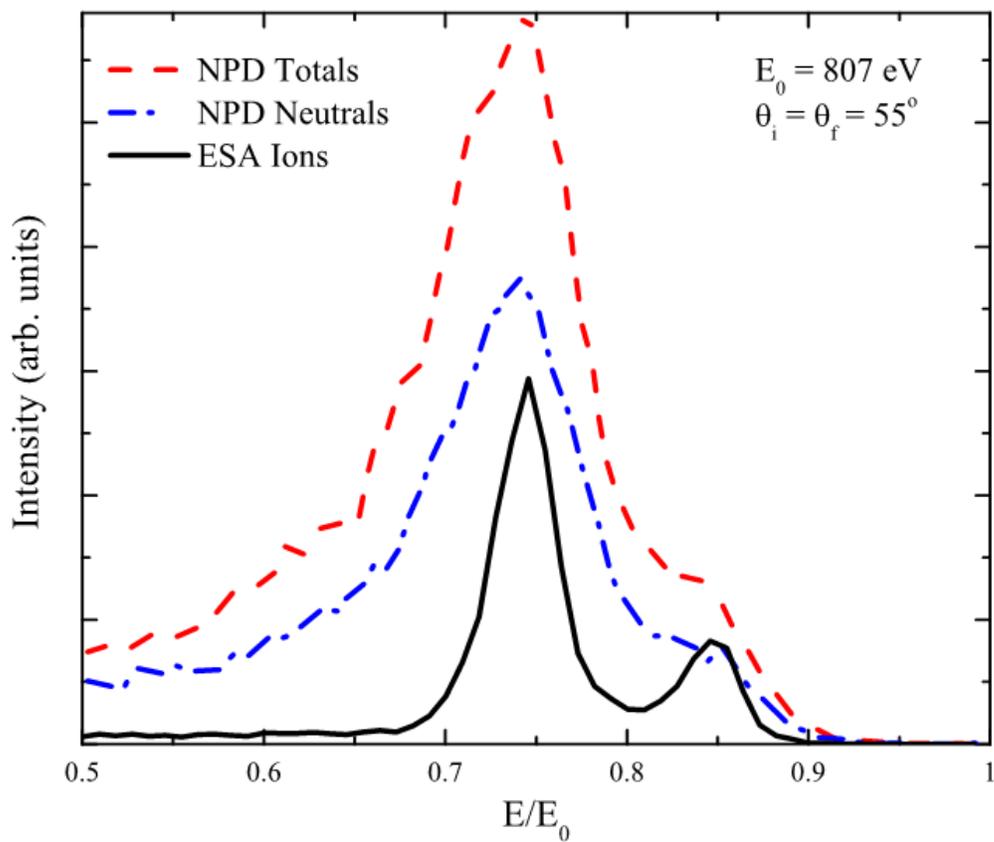

FIG. 2: A representative NPD spectrum showing both the scattered totals and neutrals intensity measured for $Na^+$ incident on Ag(001). The neutralization probability, P0, obtained for this spectrum was approximately 31%. The corresponding scattered spectrum obtained with the ESA is also shown for comparison.

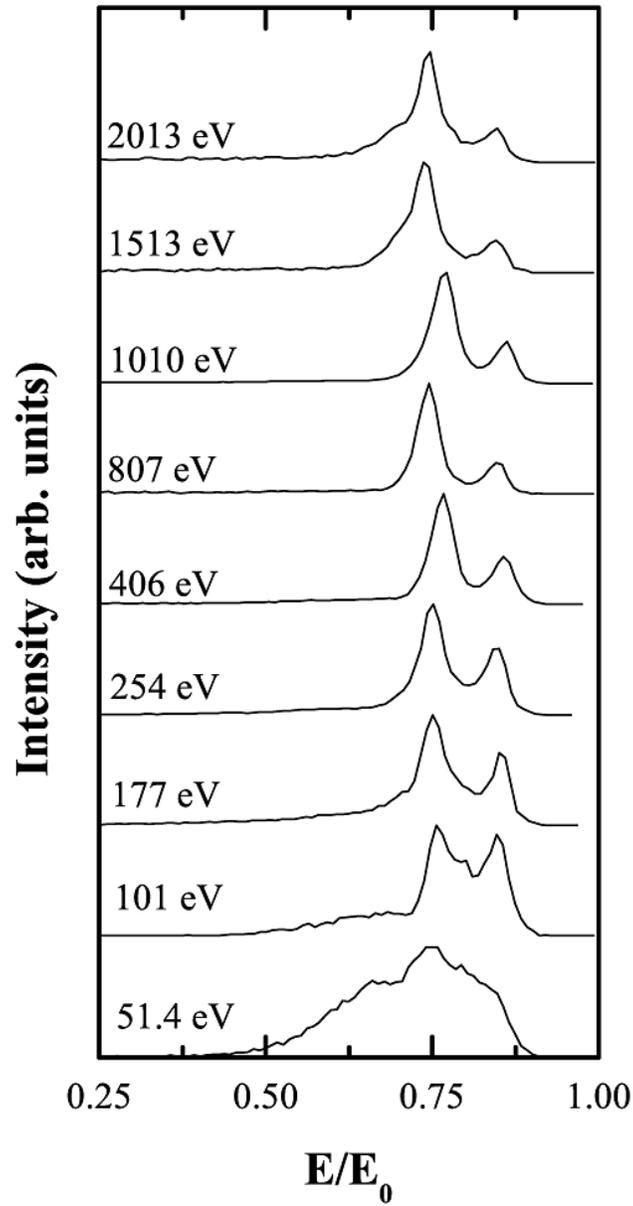

FIG. 3: Energy loss spectra for specular scattering at 55° over a range of incident beam energies. For comparison the energy axis has been scaled by the incident beam energy. The two most prominent peaks in the spectra are associated with the QS and QD scattering trajectories.

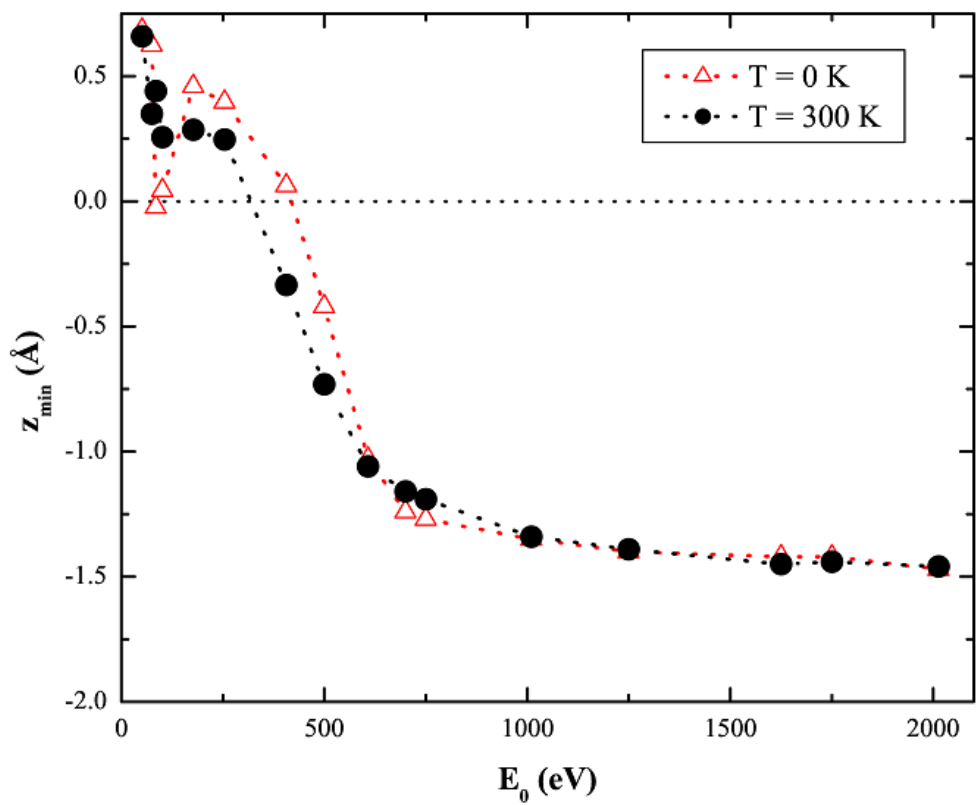

FIG. 4: (Color online) Simulated data of the penetration depth as a function of incident beam energy. The simulation was performed at 0 K and 300 K. The simulations show that the penetration depth is sensitive to thermal displacements of the surface atoms. A line has been drawn between data points to guide the eye. The dotted line at $z_{min}$ = 0.0 Å corresponds to the center of the Ag surface atoms. For these results the calculated variance is smaller than the symbol size at each beam energy.

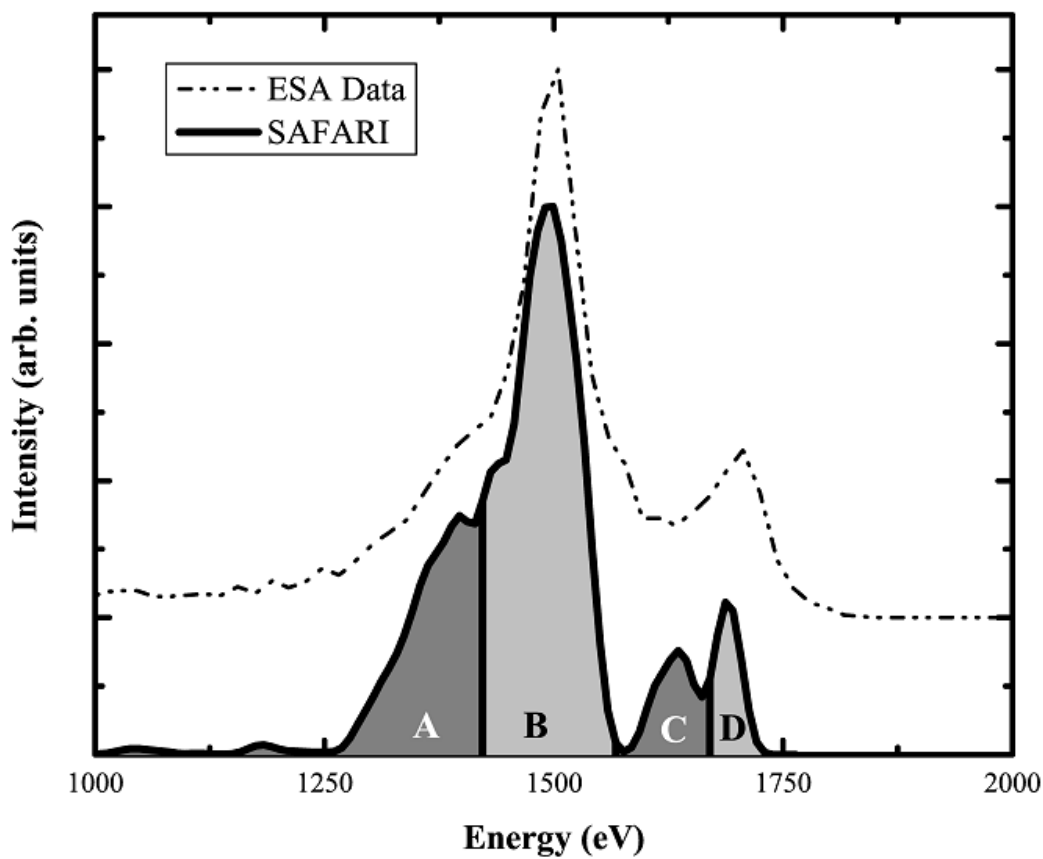

FIG. 5: A comparison of scattered spectra obtained with the ESA (dashed-dotted line) and with the trajectory simulation SAFARI (solid line) for Na$^+$ incident on Ag(001) along the [110] direction at $\theta_i= \theta_f$ =55° and $E_0$ = 2013 eV. The simulated spectrum has been divided up into four energetic regions (A-D) to facilitate discussion of the contributing trajectory types for each region. The ESA data have been offset for clarity. A detailed peak analysis indicates an agreement (within 1%) between these data and the simulated SAFARI spectrum.

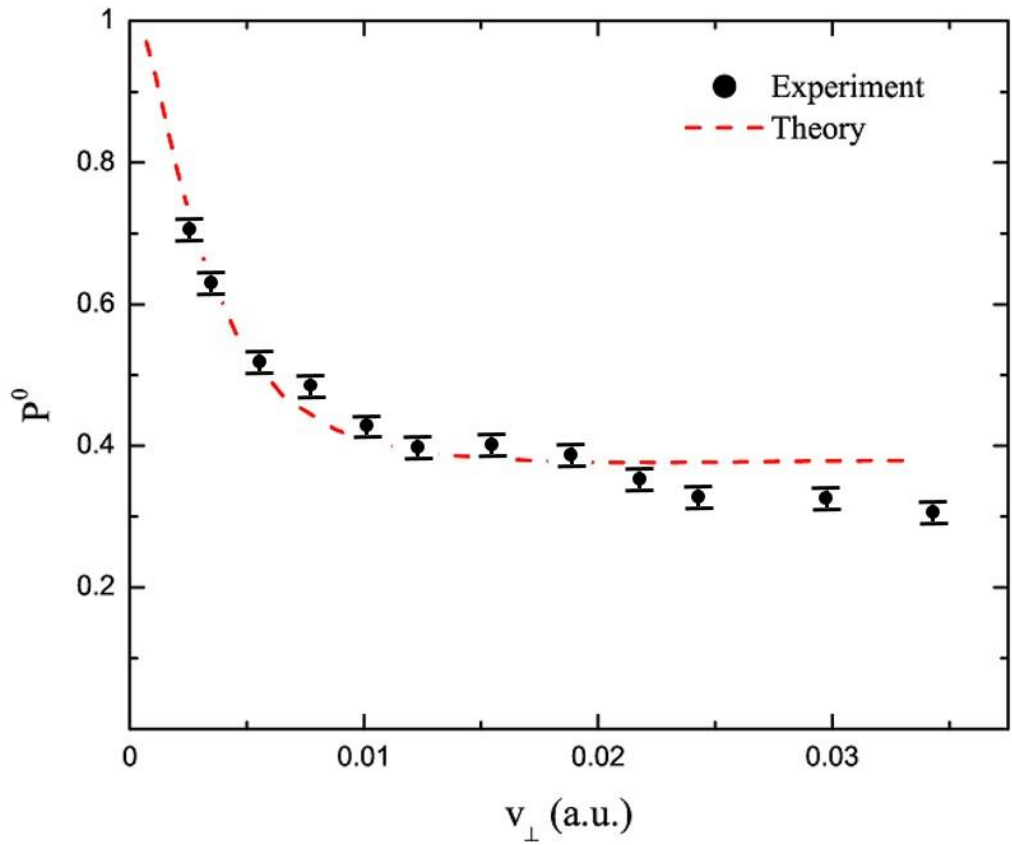

FIG. 6: (Color online) A comparison between the quantum mechanical model INDEP and the experimental data for the neutralization probability as a function of the incident ion velocity. Each data point in the plot represents the scattered neutral to ion flux ratio obtained by methods described in the text. The trend with incident velocity is consistent with a resonant charge transfer mechanism.